\newcommand{\be}{\begin{eqnarray}}
\newcommand{\ee}{\end{eqnarray}}
\documentclass[twocolumn,aps,showpacs]{revtex4}
\usepackage{graphics}
\begin{document}
\title{$\rho$ propagation and dilepton production at finite pion density
and temperature} 
\author{Alejandro Ayala}  
\affiliation{Instituto de Ciencias Nucleares, Universidad Nacional 
         Aut\'onoma de M\'exico, Apartado Postal 70-543, 
         M\'exico Distrito Federal 04510, M\'exico.}
\author{Javier Magnin}
\affiliation{Centro Brasileiro de Pesquisas Fisicas, Rua Dr. Xavier
             Sigaud 150 - URCA CEP 22290-180 Rio de Janeiro Brazil.}
\begin{abstract}

We study the propagation properties of the $\rho$ vector in a dense
and hot pion medium. We introduce a finite value of the chemical
potential associated to a conserved pion number and argue that such
description is valid during the hadronic phase of a relativistic
heavy-ion collision, between chemical and thermal freeze-out, where the
strong interaction drives pion number to a fixed value. By invoking
vector dominance and $\rho$ saturation, we also study the finite pion
density effects into the low mass dilepton production rate. We find
that the distribution moderately widens and the position of the peak
shifts toward larger values of the pair invariant mass, at the same
time that the height of the peak decreases when the value of the
chemical potential grows. We conclude by arguing that for the
description of the dilepton spectra at ultra-relativistic energies,
such as those of RHIC and LHC, the proper treatment of the large pion
density might be a more important effect to consider than the
influence of a finite baryon density. 

\end{abstract}


\maketitle

\section{Introduction}

One of the most salient features of the low-mass dilepton spectra in
relativistic nucleus-nucleus collisions, from BEVALAC/SIS to SPS
energies, is the enhancement in the production yields for invariant
masses between 0.2 and 1 GeV, as compared to proton induced
reactions~\cite{experiments}. Since dilepton final states are mediated
by electromagnetic currents and these in turn are connected to vector
mesons, dilepton pairs represent a prime tool to study the evolution
of the dense and hot hadronic region formed in this kind of
collisions. For low invariant masses, the vector mesons involved are the
$\rho$, $\omega$ and $\phi$. Among these, $\rho$ plays an special
role given that its lifetime is smaller than the expected lifetime
of the interacting region and thus is able to probe different
stages during the collision of heavy systems.  

The favored explanations, able to account for a great
deal of the features of the measured low-mass dilepton spectra can be
divided in two categories: the {\it dropping} $\rho$ mass and the {\it
melting} of resonances scenarios~\cite{Rapp}. The first of these,
connected to the Brwon-Rho scaling conjecture~\cite{Brown} and the
decrease of the quiral quark condensate with temperature and baryon
density, states that the in-medium $\rho$ mass will sweep the entire
low invariant mass region as the system cools down from its initially
hot and dense state toward freeze-out. The second
scenario~\cite{Dominguez, Asakawa} states that the in-medium spectral
densities of the $\rho$ and its chiral partner, the $a_1$ (1260)
become broad and structureless, merging into a flat continuum as the
system approaches chiral symmetry restoration. 

An important ingredient for the success of the two above mentioned
scenarios is the presence of finite baryon density effects (see for
example Ref.~\cite{Chen}). In the
first case, baryons act as the source of strongly attractive scalar
fields. In the second case, interactions of the vector mesons with
baryonic resonances, characterized by large coupling constants,
overwhelm the relative strength of the interactions of $\rho$'s with
pions, despite the fact that at SPS energies, the relative abundance
of the latter is five times that of the former.

Nevertheless, at RHIC and moreover at LHC energies, the central
rapidity region is expected to become baryon free with the relative
abundance of pions being larger than at SPS energies. Consequently, at
ultra-relativistic energies, it is important to include in the
calculation of the dilepton spectrum the proper treatment of the
large pion density, particularly during the hadronic phase of
the collision. 

Recall that strictly speaking, pion number is not a conserved
quantity and that pion decay is driven by all the relevant
interactions, namely, strong, weak and electromagnetic.
However, the characteristic time for electromagnetic
and weak pion number-changing reactions, is very large compared to the
lifetime of the system created in relativistic heavy-ion collisions
and therefore, these processes are of no relevance for the propagation
properties of pions within the lifetime of the collision. As for the
case of strong processes, it is by now accepted that they drive pion
number toward chemical freeze-out at a temperature considerably higher
than the thermal freeze-out temperature and therefore, that
from chemical to thermal freeze-out, the pion system evolves with the
pion abundance held fixed~\cite{Bebie, Braun-Munzinger}. Under these
circumstances, it is possible to ascribe to the pion density a
chemical potential and consider the pion number as
conserved~\cite{Hung,Chungsik}. In this context, the role of a finite
pion chemical potential into a hadronic equation of state has been recently
investigated in Refs.~\cite{Teaney}. The effects of a finite isospin
chemical potential on the pion mass have also been recently studied in
Refs.~\cite{Loewe}. 

The description of hadronic degrees of freedom belongs to the realm of
nonperturbative phenomena and therefore has necessarily to rely on
effective approaches that implement the dynamical symmetries of
QCD. In a series of recent papers~\cite{Ayala, Ayala2} it has been
shown that the linear sigma model can be used as one of such effective
approaches to describe the pion propagation properties within a pion
medium at energies, temperatures and densities small compared to
the sigma mass. The sigma degree of freedom can be integrated out in a
systematic expansion to obtain an effective theory of like-isospin
pions interacting among themselves through an effective quartic
term with coupling $\alpha = 6(m_\pi^2/2f_\pi^2)$, where $m_\pi$ and
$f_\pi$ are the vacuum pion mass and decay constant, respectively.

In this paper we extend the use of such effective description to
study the interaction of pions with the $\rho$ vector, paying
particular attention to the effects that a finite pion density
introduce on the propagation properties of $\rho$ at finite
temperature. We find the finite density and temperature modifications
to the $\rho$ mass, width and dispersion relation. By invoking vector
dominance, we also study the effects that these modifications
introduce on the production of $e^+$ $e^-$ pairs near the $\rho$ peak.

\begin{figure}[t!] 
{\centering
{\includegraphics{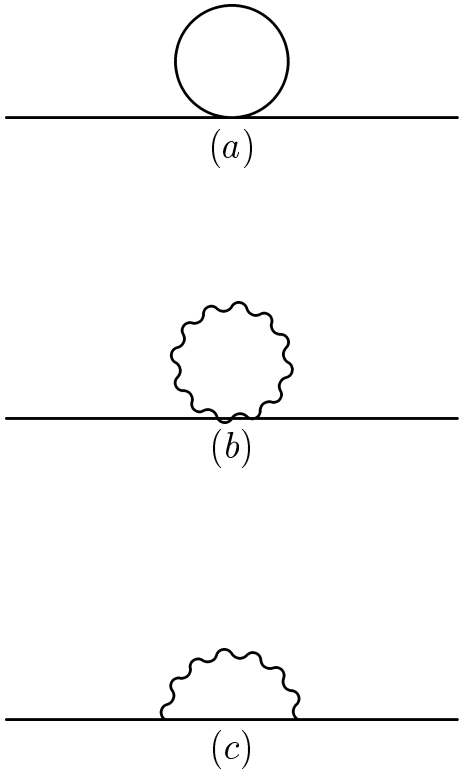}}
\par}
\vspace{-3.0in}
\caption{Feynman diagrams representing the pion self-energy at one
loop. The wavy lines represent the $\rho$ whereas the
solid lines represent the pion. In the approximation where $m_\rho \gg
m_\pi, T$, only diagram $(a)$ contributes.} 
\end{figure}

The work is organized as follows: In Sec.~\ref{secII}, we introduce
the $\rho$ in the description by {\it gauging} the original effective
Lagrangian. The theory thus obtained closely resembles scalar
electrodynamics. We introduce a chemical potential associated to a
conserved number of pions and construct the modifications to the pion
propagator and $\pi$-$\rho$ vertex at finite density. In
Sec.~\ref{secIII}, we use this effective Lagrangian to compute
the pion self-energy and in Sec.~\ref{secIV} the $\rho$ self-energy and
from it, the modifications to its mass, width 
and dispersion curve at finite density and temperature. In
Sec.~\ref{secV}, we compute the dilepton rate assuming vector
dominance. We finally summarize and conclude in Sec.~\ref{secV}.

\section{Effective Lagrangian}\label{secII}

In Refs.~\cite{Ayala} it has been shown that starting from the linear
sigma model Lagrangian, including only meson degrees of freedom and
working in the kinematical regime where the pion momentum, mass and
temperature are small compared to the sigma mass, the two-loop pion
self energy can be formally obtained by means of the effective
Lagrangian  
\be
   {\mathcal L}=\frac{1}{2}\left(\partial^\mu{\mathbf{\phi}}\right)^2
   -\frac{1}{2}m_\pi^2{\mathbf{\phi}}^2 -\frac{\alpha}{4!}
   {\mathbf{\phi}}^4\, ,
   \label{effLag}
\ee
\begin{figure}[t!] 
{\centering
{\includegraphics{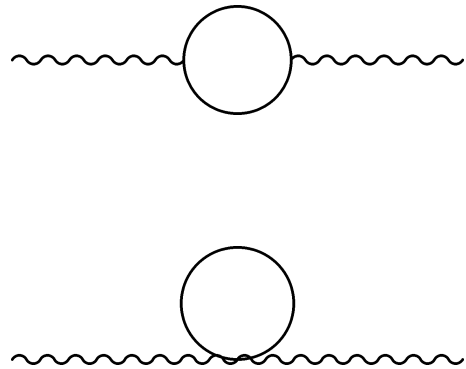}}
\par}
\vspace{-4.8in}
\caption{Feynman diagrams representing the $\rho$ self-energy at one
loop. The wavy lines represent the $\rho$ whereas the solid lines
represent the pion.} 
\end{figure}
where $\alpha=6(m_\pi^2/2f_\pi^2)$ and the factor $6$ comes from
considering the interaction of like-isospin pions in the vertex
\be
   i\Gamma_4^{ijkl}=-2i\left(\frac{m_\pi^2}{2f_\pi^2}\right)
   \left(\delta^{ij}\delta^{kl}+\delta^{ik}\delta^{jl}+\delta^{il}\delta^{jk}
   \right)\, .
   \label{vertmod}
\ee
In essence, the theory thus constructed and summarized by the
effective Lagrangian in Eq.~(\ref{effLag}) can be thought of as a theory
for the effective coupling $\alpha$ that encodes the dynamics of low
energy pion interactions. It can also be checked that
Eq.~(\ref{effLag}) reproduces the leading order modification to the
pion mass at finite temperature obtained from chiral perturbation
theory~\cite{Gasser}. 

In order to introduce the $\rho$ field, we {\it gauge} the
theory~\cite{Gale, Pisarski} described by the Lagrangian in
Eq.~(\ref{effLag}) replacing the derivative $\partial^\mu$ by the
covariant derivative $D^\mu$ given by 
\be
   \partial^\mu \phi\rightarrow D^\mu\phi = (\partial^\mu
   -ig\rho^\mu)\phi\, ,
   \label{covder}
\ee
where we have introduced the $\pi$-$\rho$ coupling constant $g$. Also,
by introducing the mass term and kinetic energy for the $\rho$ field,
the Lagrangian in Eq.~(\ref{effLag}) becomes
\be
   {\mathcal L}\rightarrow {\mathcal L}' &=& 
   \frac{1}{2}(D^\mu\phi)^2
   -\frac{1}{2}m_\pi^2\phi^2 -\frac{\alpha}{4!}
   \phi^4\nonumber\\
   &+&\frac{1}{2}m_\rho^2\rho^\mu\rho_\mu
   -\frac{1}{4}\rho_{\mu\nu}\rho^{\mu\nu}\, .
   \label{effLagprim}
\ee
\begin{figure}[t]
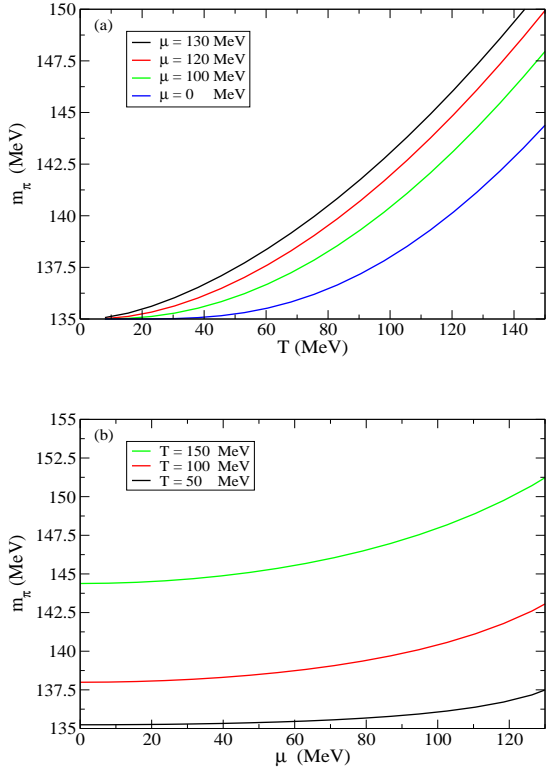
 
{\centering
\resizebox*{0.4\textwidth}{0.2\textheight}{\includegraphics{amfig3a.eps}}
\par}
\vspace{0.7cm}
{\centering 
\resizebox*{0.4\textwidth}{0.2\textheight}{\includegraphics{amfig3b.eps}}
\par}
\caption{Thermal pion mass as a function of $(a)$ $T$ for
different values of $\mu$ ranging 
from $\mu=0$ to $\mu=130$ MeV and as a function of $(b)$ $\mu$ for
different values of $T$ ranging from $T=50$ to $T=150$ MeV.}
\end{figure}
Let us now pause briefly to describe the formalism that allows us to
introduce a finite chemical potential associated to a conserved pion
number. To this end, let us further modify the Lagrangian in
Eq.~(\ref{effLagprim}), writing it in terms of a complex scalar field
and regarding $\phi$ and $\phi^*$ as independent fields
\be
   {\mathcal L}'\rightarrow {\mathcal L}'' &=& 
   (D_\mu\phi)(D^\mu\phi^*)
   -m_\pi^2\phi\phi^* -\frac{\alpha}{4}
   (\phi\phi^*)^2\nonumber\\
   &+&\frac{1}{2}m_\rho^2\rho^\mu\rho_\mu
   -\frac{1}{4}\rho_{\mu\nu}\rho^{\mu\nu}\, .
   \label{effLagdoubleprim}
\ee
\begin{figure}[t]
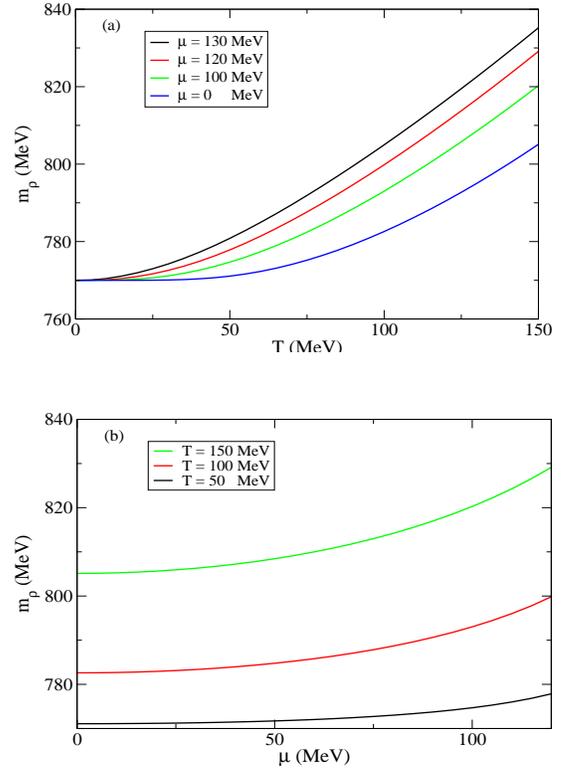
 
{\centering
\resizebox*{0.4\textwidth}{0.2\textheight}{\includegraphics{amfig4a.eps}}
\par}
\vspace{0.7cm}
{\centering 
\resizebox*{0.4\textwidth}{0.2\textheight}{\includegraphics{amfig4b.eps}}
\par}
\caption{Thermal $\rho$ mass as a function of $(a)$ $T$ for
different values of $\mu$ ranging 
from $\mu=0$ to $\mu=130$ MeV and as a function of $(b)$ $\mu$ for
different values of $T$ ranging from $T=50$ to $T=150$ MeV.} 
\end{figure}
The effective Lagrangian in Eq.~(\ref{effLagdoubleprim}) resembles
that of scalar electrodynamics with the photon field replaced by the
massive $\rho$ field. Invariance under global phase transformations 
\be
   \phi\rightarrow\phi' = e^{-i\lambda}\phi
   \label{globalphase}
\ee
leads to the conserved current
\be
   J^\mu = i(\phi^*\partial^\mu\phi - \phi\ \partial^\mu\phi^*) -
   2g\rho^\mu\phi^*\phi\, ,
   \label{conscurr}
\ee
and to the conserved charge $N$, that can be identified with the particle
number, given by
\be
   N=i\int d^3x(\phi^*\partial^0\phi - \phi\ \partial^0\phi^* +
   2ig\rho^0\phi^*\phi)\, .
   \label{conscharge}
\ee 
\begin{figure}[t]
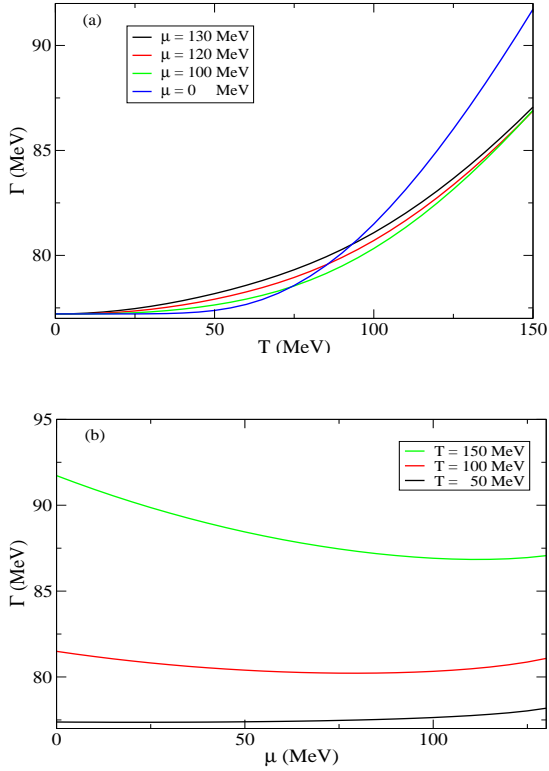
 
{\centering
\resizebox*{0.4\textwidth}{0.2\textheight}{\includegraphics{amfig5a.eps}}
\par}
\vspace{0.7cm}
{\centering 
\resizebox*{0.4\textwidth}{0.2\textheight}{\includegraphics{amfig5b.eps}}
\par}
\caption{Thermal $\rho$ half width as a function of $(a)$ $T$ for
different values of $\mu$ ranging 
from $\mu=0$ to $\mu=130$ MeV and as a function of $(b)$ $\mu$ for
different values of $T$ ranging from $T=50$ to $T=150$ MeV.} 
\end{figure}
Since $N$ is a conserved charge, there exists a chemical
potential $\mu$ conjugate to $N$ so that the grand partition function
is
\be
   Z(\beta ,\mu)={\mbox {Tr}}\ e^{-\beta(H-\mu N)}\, ,
   \label{partfunc}
\ee
where $\beta = 1/T$ is the inverse of the temperature $T$. From now
on, let us work explicitly in the imaginary-time formalism of thermal
field theory. It is straightforward to write a path integral
representation in Euclidean space of the grand partition function
$Z(\beta ,\mu)$. Care has to be taken by going through the Hamiltonian
form since $N$ depends on the time derivative of $\phi$ and
$\phi^*$. After integration over the conjugate fields
$\pi^*=\partial{\mathcal L}''/\partial(\partial^0\phi^*)$ and
$\pi=\partial{\mathcal L}''/\partial(\partial^0\phi)$ one can check
that in the exponent, there appear the combinations
\be
    -\phi^*\left(\frac{\partial^2}{\partial\tau^2} -
    2\mu\frac{\partial}{\partial\tau} + \mu^2 -
    m_\pi^2\right)\phi\, ,
    \label{comb1}
\ee
\be
    ig\rho^0\left(\phi^*\frac{\partial\phi}{\partial\tau} -
    \phi\frac{\partial\phi^*}{\partial\tau}\right) -
    2ig\mu\rho^0\phi^*\phi\, ,
    \label{comb2}
\ee
where $\tau$ is the Euclidean time. Going to frequency space the
Euclidean time derivatives get replaced by
\be
   \frac{\partial\phi}{\partial\tau}&\rightarrow&
   -i\omega_n\phi\nonumber\\ 
   \frac{\partial\phi^*}{\partial\tau}&\rightarrow&
   i{\omega}_{n'}\phi^*\, , 
   \label{replace}
\ee
\begin{figure}[t!] 
{\centering
\resizebox*{0.44\textwidth}{0.22\textheight}{\includegraphics{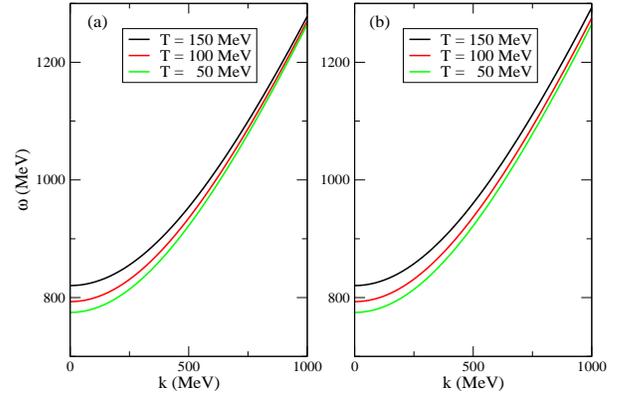}}
\par}
\caption{$\rho$ dispersion relation for $(a)$ longitudinal and $(b)$
transverse modes for $\mu=100$ MeV for
different values of $T$ ranging from $T=50$ to $T=150$ MeV.} 
\end{figure}
where the periodicity of the fields in the interval $ 0\leq \tau \leq
\beta$ makes the (Matsubara) frequencies $\omega_n$ and ${\omega}_{n'}$
be discrete and integer multiples of $2\pi T$. The combination in
Eq.~(\ref{comb1}) translates into a modification of the Matsubara pion
propagator which now reads as (hereafter capital letters are used to
denote four-vectors whereas lower case letters are used do denote the
components)
\be
   \Delta(i\omega_n,p;\mu)=\frac{1}{-(i\omega_n +\mu)^2 + p^2
   +m_\pi^2}\, ,
   \label{prop}
\ee
whereas the combination in Eq.~(\ref{comb2}) goes into a modification
of the $\pi$-$\rho$ vertex which now becomes
\be
   \Gamma_{\pi\rho}(P_\mu, {P'}_\mu;\mu)&=&
   -ig\{[-(i\omega_n + \mu),{\mathbf p}]\nonumber\\
   &+& [-(i{\omega}_{n'} + \mu),{\mathbf p'}]\}\, , 
   \label{vert}
\ee
The final outcome is that the introduction of a finite chemical
potential translates into the substitution
\be
   i\omega_n\rightarrow i\omega_n + \mu
   \label{subst}
\ee
both when this frequency appears in internal pion lines either in the
Matsubara propagator or in the the $\pi$-$\rho$ vertex. Notice that
these substitutions agree with the weel known result for the
periodicity of the Matsubara propagator in the mixed representation
given by
\be
   \Delta (\beta-\tau,E;\mu)=\Delta (\tau,E;-\mu)\, .
   \label{Matsbound}
\ee

\section{$\pi$ self-energy}\label{secIII}

The effective Lagrangian in Eq.~(\ref{effLagdoubleprim}) describes the
interactions between pions and $\rho$'s as well as among pions
themselves. The one-loop diagrams for the pion and $\rho$ 
self-energies are depicted in Figs.~1 and~2. For the pion self-energy, the
diagrams contain terms with internal $\rho$ propagators. Notice that
since $m_\rho \gg m_\pi,\ T$, these diagrams are strongly suppressed
at finite temperature compared to those made out exclusively of
internal pion lines and can thus be neglected. This approximation goes
along the lines of the reasoning used in Refs.~\cite{Ayala} where the
heavy internal sigma modes were systematically {\it pinched} to
construct the effective vertices and propagators leading to the
effective Lagrangian in Eq.~(\ref{effLag}). In this scheme, the pion
and $\rho$ self-energies decouple and the former gets modified only by
the diagram in Fig.~1a leading to a momentum independent correction of
the pion mass. We can thus take one step further going beyond naive
perturbation theory and adopt a resummation scheme for the pion
self-energy to look for the modification of the pion mass beyond
leading order. This can be implemented by writing the pion self-energy
explicitly as 
\be
   \Pi_0=\frac{\alpha}{2}T\sum_n\int\frac{d^3k}{(2\pi)^3}
   \frac{1}{K^2+m_\pi^2+\Pi_0}\, ,
   \label{pi0}
\ee

Equation~(\ref{pi0}) represents a self consistent relation for the
temperature and density dependent quantity $\Pi_0$. This is the well known
resummation for the {\it superdaisy} diagrams which constitute the dominant
contribution in the large-$N$ expansion~\cite{Dolan} of the Lagrangian in
Eq.~(\ref{effLag}). The solution to Eq.~(\ref{pi0}) is given by the
transcendental equation 
\be
   \Pi_0&=&\left(\frac{\alpha T}{4\pi^2}\right)\sqrt{m_\pi^2 + \Pi_0}
   \nonumber\\
   &\times&\sum_{n=1}^\infty K_1\left(\frac{n\sqrt{m_\pi^2 + \Pi_0}}{T}\right)
   \frac{\cosh (n\mu /T)}{n}\, .
   \label{solpi0}
\ee
Figure~3 shows the behavior of the pion thermal mass
$\tilde{m}_\pi=\sqrt{m_\pi^2 + \Pi_0}$ as a function of $(a)$ $T$ for
different values of $\mu$ and $(b)$ as a function of $\mu$ for
different values of $T$. We use the values $m_\pi=135$ MeV, $f_\pi=93$
MeV. From Fig.~3, we notice that $\tilde{m}_\pi$ grows monotonically
with both $T$ and $\mu$. 

\begin{figure}[t!] 
{\centering
\resizebox*{0.44\textwidth}{0.22\textheight}{\includegraphics{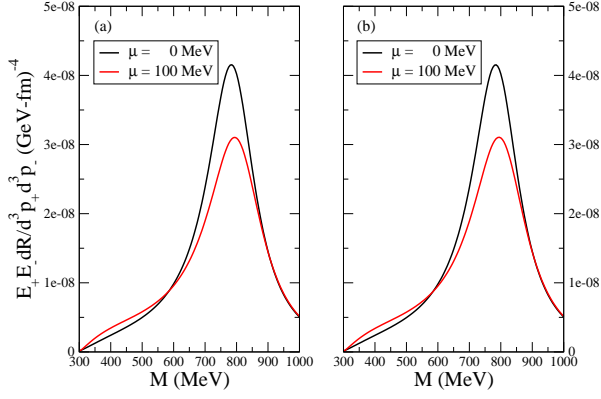}}
\par}
\caption{Dilepton production rates for $(a)$ longitudinal and $(b)$
transverse modes for $T=150$ MeV, $k=50$ MeV and $\mu=0, 100$ MeV.} 
\end{figure}

\section{$\rho$ self-energy}\label{secIV}

The explicit expression for the one-loop $\rho$ self-energy depicted in
Fig.~2 is given by
\be
   \Pi^{\mu\nu}&=&-g^2T\sum_n\int\frac{d^3p}{(2\pi)^3}
   \frac{(2P^\mu-K^\mu)(2P^\nu-K^\nu)}
   {(P^2+\tilde{m}_\pi^2)[(K-P)^2+\tilde{m}_\pi^2]}\nonumber\\
   &+&\delta^{\mu\nu}g^2T\sum_n\int\frac{d^3p}{(2\pi)^3}
   \frac{1}{P^2+\tilde{m}_\pi^2}\, ,
   \label{rhoself}
\ee
where, according to the discussion in Sec.~\ref{secII}, the internal
pion momentum $P^\mu$ and the $\rho$ momentum $K^\mu$ are 
\be
   P^\mu&=&[-(i\omega_n +\mu), {\mathbf p}]\nonumber\\
   K^\mu&=&(-i\omega, {\mathbf k})\, .
   \label{vecdef}
\ee  
Equation~(\ref{rhoself}) contains vacuum and matter contributions. It is
well known that the infinities coming from the vacuum pieces can be
reabsorbed into the redefinition of the bare masses and
couplings~\cite{Gale}. For what follows, we will concentrate on
the matter contributions.

\begin{figure}[t!] 
{\centering
\resizebox*{0.44\textwidth}{0.22\textheight}{\includegraphics{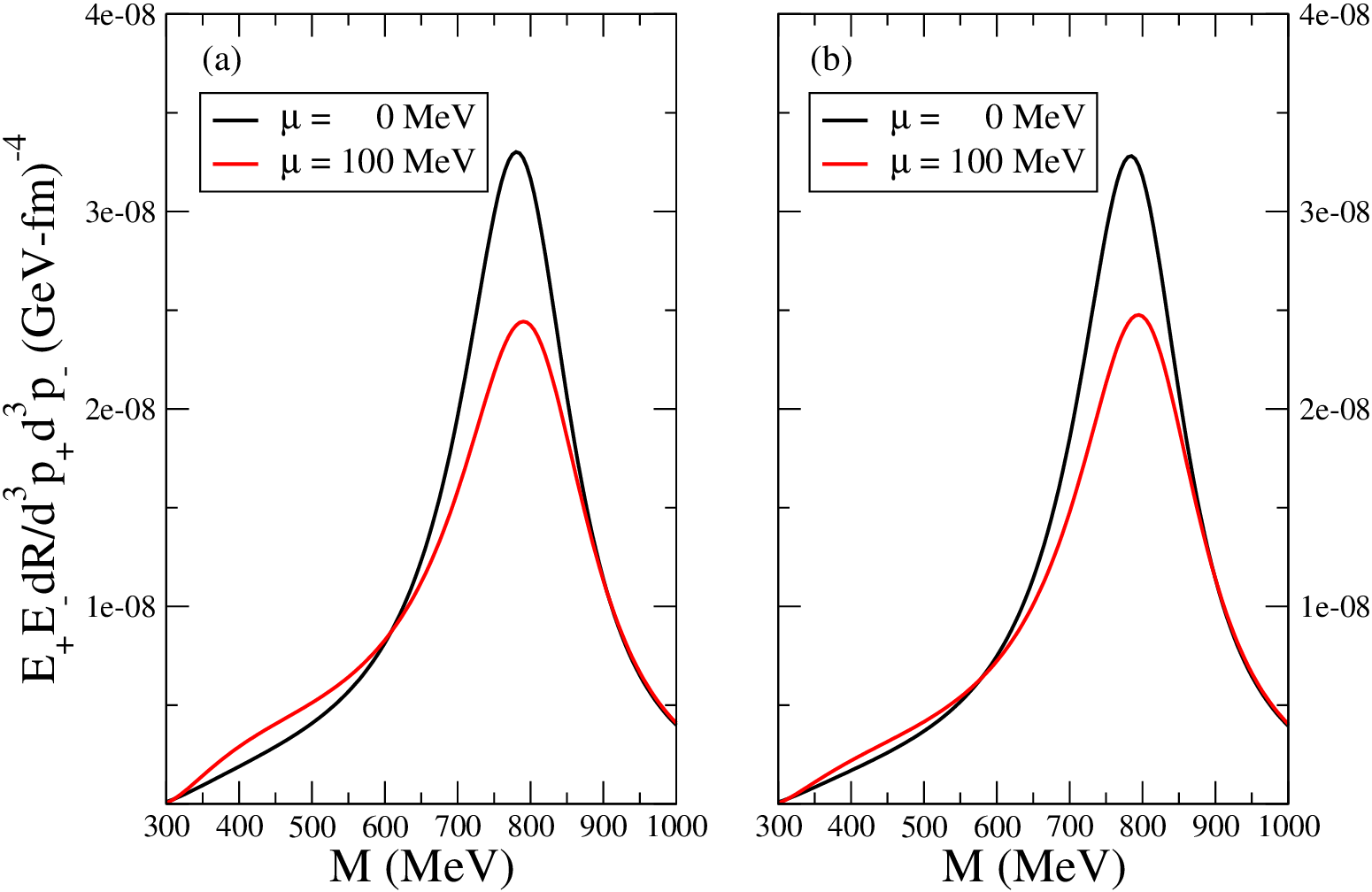}}
\par}
\caption{Dilepton production rates for $(a)$ longitudinal and $(b)$
transverse modes for $T=150$ MeV, $k=250$ MeV and $\mu=0, 100$ MeV.} 
\end{figure}

For a massive vector field, the tensor structure of its self-energy
can be written in terms of the longitudinal $P_L^{\mu\nu}$ and
transverse $P_T^{\mu\nu}$ projection tensors
\be
   \Pi^{\mu\nu}=F(K)P_L^{\mu\nu}+G(K)P_T^{\mu\nu}\, .
   \label{proj}
\ee
In Minkowski space, $P_{T,L}^{\mu\nu}$ are given by
\be
   P_T^{00}&=&P_T^{0i}=P_T^{i0}=0\nonumber\\
   P_T^{ij}&=&\delta^{ij}-k^ik^j/k^2\nonumber\\
   P_L^{\mu\nu}&=&-g^{\mu\nu}+K^\mu K^\nu/K^2 - P_T^{\mu\nu}\, .
   \label{projLT}
\ee
From the relation between the self-energy, and the full $D^{\mu\nu}$
and bare $D^{\mu\nu}_0$ $\rho$ propagators, we have, also in Minkowski
space 
\be
   -iD^{\mu\nu}=\frac{P_L^{\mu\nu}}{K^2-m_\rho^2-F}+
   \frac{P_T^{\mu\nu}}{K^2-m_\rho^2-G} + 
   \frac{K^\mu K^\nu}{m_\rho^2K^2}.
   \label{proprho}
\ee
In order to identify the coefficients $F$ and $G$ of the longitudinal
and transverse projectors, we take ${\mathbf k}$ along the
$z$-axis. Thus, in Minkowski space, their expressions are
\be
   F(K)&=&-\frac{K^2}{k_0k}\Pi^{03}\nonumber\\
   G(K)&=&\Pi^{11}\, ,
   \label{FandG}
\ee
where $\Pi^{03}$ and $\Pi^{11}$ are obtained from Eq.~(\ref{rhoself})
with the analytical continuation 
\be
   i\omega\rightarrow k_0 + i\epsilon\, ,
   \label{analcont}
\ee
that give the retarded functions and that can be performed after carrying
out the sum over the Matsubara frequencies. 

The sum over frequencies can be obtained by standard
techniques~\cite{Kapusta, LeBellac}. Considering the effects of a
finite chemical potential, the expressions of interest are
\begin{widetext}
\be
   T\sum_n \Delta(i\omega_n,p)\Delta(i(\omega-\omega_n),|{\mathbf
   k}-{\mathbf p}|)&=&-
   \sum_{s_1s_2=\pm}\frac{s_1s_2}{4E_pE_{|{\mathbf k}-{\mathbf p}|}}
   \frac{[1+n(s_1E_p-\mu)+n(s_2E_{|{\mathbf k}-{\mathbf p}|}+\mu)]}
   {(i\omega - s_1E_p - s_2E_{|{\mathbf k}-{\mathbf p}|})}\nonumber\\
   T\sum_n i\omega_n\Delta(i\omega_n,p)\Delta(i(\omega-\omega_n),|{\mathbf
   k}-{\mathbf p}|)&=&-
   \sum_{s_1s_2=\pm}\frac{s_1s_2(s_1E_p-\mu)}
   {4E_pE_{|{\mathbf k}-{\mathbf p}|}}
   \frac{[1+n(s_1E_p-\mu)+n(s_2E_{|{\mathbf k}-{\mathbf p}|}+\mu)]}
   {(i\omega - s_1E_p - s_2E_{|{\mathbf k}-{\mathbf p}|})}\, ,
   \label{sumsexp}
\ee
\end{widetext}
where 
\be
   n(x)=\frac{1}{e^{\beta |x|}-1}\, ,
   \label{be}
\ee
is the Bose-Einstein distribution,
$E_p=\sqrt{p^2+\tilde{m}_\pi^2}$, $E_{|{\mathbf k}-{\mathbf
p}|}=\sqrt{({\mathbf k}-{\mathbf p})^2+\tilde{m}_\pi^2}$ and the
function $\Delta$ is defined in Eq.~(\ref{prop}).

Using Eq.~(\ref{sumsexp}) into Eq.~(\ref{rhoself}) and by means of the
analytical continuation in Eq.~(\ref{analcont}), the real and
imaginary parts of $\Pi^{03}$ and $\Pi^{11}$ are given by
\begin{widetext}
\be
   {\mbox {Re}}\Pi^{03}&=&\frac{g^2}{8\pi^2}\int_0^\infty
   dp\left(\frac{p^2}{E_p}\right)\left[n(E_p-\mu)+n(E_p+\mu)\right]
   \left[k_0(A_-+A_+)-2E_p(A_--A_+)\right]\nonumber\\
   {\mbox {Im}}\Pi^{03}&=&\frac{g^2}{16\pi}\left(\frac{k_0}{k^2}\right)
   \Big\{\int_{\frac{k_0-ak}{2}}^{\frac{k_0+ak}{2}}
   dE_p(2E_p-k_0)^2[1+2n(E_p-k_0-\mu)]\theta (k_0)\nonumber\\
   &-&
   \int_{\frac{-k_0-ak}{2}}^{\frac{-k_0+ak}{2}}
   dE_p(2E_p+k_0)^2[1+2n(E_p+k_0+\mu)]\theta (-k_0)\Big\}\nonumber\\
   {\mbox {Re}}\Pi^{11}&=&\frac{g^2}{8\pi^2}\int_0^\infty
   dp\left(\frac{p^2}{E_p}\right)\left[n(E_p-\mu)+n(E_p+\mu)\right]
   \left[\frac{2(k_0^2+k^2)}{k^2}+(B_-+B_+)\right]\nonumber\\
   {\mbox {Im}}\Pi^{11}&=&-\frac{g^2}{32\pi}\left(\frac{1}{k^3}\right)
   \Big\{\int_{\frac{k_0-ak}{2}}^{\frac{k_0+ak}{2}}
   dE_p[4p^2k^2-(k^2-k_0^2+2k_0E_p)^2]
   [1+2n(E_p-k_0-\mu)]\theta (k_0)\nonumber\\
   &-&
   \int_{\frac{-k_0-ak}{2}}^{\frac{-k_0+ak}{2}}
   dE_p[4p^2k^2-(k^2-k_0^2-2k_0E_p)^2]
   [1+2n(E_p+k_0+\mu)]\theta (-k_0)\Big\}\nonumber\\
   A_\pm&=&\frac{1}{2k^2p}\left(4kp + [k_0^2\pm2k_0E_p]
   \ln\left|\frac{k_0^2-k^2\pm2k_0E_p-2kp}{k_0^2-k^2\pm2k_0E_p+2kp}
   \right|\right)\nonumber\\
   B_\pm&=&\left(\frac{(k_0^2-k^2\pm 2k_0E_p)^2-4k^2p^2}{4k^3p}\right)
   \ln\left|\frac{k_0^2-k^2\pm2k_0E_p-2kp}{k_0^2-k^2\pm2k_0E_p+2kp}\right|\, ,
   \label{ABpm}
\ee
\end{widetext}
where the function $a$ is defined as
\be
   a=\sqrt{1-\frac{4\tilde{m}_\pi^2}{(k_0^2-k^2)}}\, .
   \label{funca}
\ee
It is easy to check that in the limit $\mu\rightarrow 0$,
Eqs.~(\ref{ABpm}) reduce to the corresponding expressions found in
Ref.~\cite{Gale}. 

Figure~4 shows the behavior of the $\rho$ thermal mass obtained as the
solution for $k_0$ of either
\be
   k_0^2- m_\rho^2 - {\mbox {Re}}\left\{\begin{array}{c}
                                 F(k_0,k=0)\\
                                 G(k_0,k=0)
                                 \end{array}\right\}=0\, ,
   \label{disprho}
\ee
as a function of $(a)$ $T$ for different values of $\mu$ and $(b)$ as
a function of $\mu$ for different values of $T$. We use the value
$g^2/4\pi=2.93$ as determined by the $\rho$ width in vacuum. For
$k=0$ there is no distinction between transverse and longitudinal
modes and thus both 
Eqs.~(\ref{disprho}) lead to the same solution. For every value of
$(T,\mu)$, the solution is computed with the corresponding value for
$\tilde{m}_\pi (T,\mu)$. From Fig.~4, we see that
the thermal $\rho$ mass grows monotonically with both $T$ and $\mu$
and that the growth is larger for larger values of $\mu$.

Figure~5 shows the behavior of the $\rho$ decay rate (or half width)
obtained from either
\be
   \Gamma = -\frac{{\mbox {Im}}\left\{\begin{array}{c}
                             F(m_\rho(T,\mu),k=0)\\
                             G(m_\rho(T,\mu),k=0)
                            \end{array}\right\}}{2m_\rho(T,\mu)}
   \label{halfwidth}
\ee
as a function of $(a)$ $T$ for different values of $\mu$ and $(b)$ as
a function of $\mu$ for different values of $T$. Again, for $k=0$
there is no distinction between 
transverse and longitudinal modes. This can be shown analytically from
the explicit expressions of ${\mbox {Im}} F$ and ${\mbox {Im}} G$ in
Eqs.~(\ref{ABpm}) which for $k=0$ yield
\be
   {\mbox {Im}} F(k_0,k=0) &=& {\mbox {Im}} G(k_0,k=0)\nonumber\\
   &=&-\left(\frac{g^2}{48\pi}\right)k_0^2
   \left(1-\frac{4\tilde{m}_\pi^2}{k_0^2}\right)^{3/2}\nonumber\\
   &\times&\left[1+2n(k_0/2+\mu)\right]\, .
   \label{IMFGk0}
\ee
For $\mu=0$, Eq.~(\ref{IMFGk0}) coincides with the corresponding
expression in Ref.~\cite{Gale}.

From Fig.~5a we see that for a given value of $\mu$ the $\rho$ width
increases monotonically with temperature. From Fig.~5b we
observe that the width reaches a minimum at a finite value of $\mu$
and that this value increases as the temperature increases. This
behavior can be understood if we recall that when the density increases
so does the pion mass and thus the phase space available for the decay
products of $\rho$ narrows. The situation changes when the density is
large enough so that the increase in the mass of the $\rho$ becomes
steeper than the growth in the pion mass, widening the phase space
available for the decay process. 

Figure~6 shows the dispersion relation for $(a)$ longitudinal and
$(b)$ transverse $\rho$ modes for different values of $T$ and $\mu=100$
MeV. The main difference between the curves in each set can be
attributed to the increase of the $\rho$  mass with both $\mu$ and $T$.

\section{Dilepton rate}\label{secV}

The electromagnetic current can be identified with the underlying quark
structure of hadrons. For invariant masses below the charm threshold,
this current can be decomposed as
\be
   j_\mu^{\ \mbox {\small {em}}}=\frac{2}{3}\bar{u}\gamma_\mu u
   -\frac{1}{3}\bar{d}\gamma_\mu d -\frac{1}{3}\bar{s}\gamma_\mu s\, .
   \label{jemquarks}
\ee
Using the $SU(3)$ quark content of hadrons, this current, being
vectorial in nature, can be in turn identified with the current
constructed out of the vector mesons $\rho$, $\omega$ and $\phi$.
\be
   j_\mu^{\ \mbox {\small {em}}}=j_\mu^\rho + j_\mu^\omega + j_\mu^\phi\, ,
   \label{jemvectors}
\ee
This is the well known conjecture named {\it vector dominance model}
(VDM). Since the pion electromagnetic form factor is is almost
totally dominated by the $\rho$ for invariant masses below 1~GeV, 
a simplified picture to study the decay of this electromagnetic
current into low mass dilepton pairs is to consider that the current
in Eq.~(\ref{jemvectors}) is totally dominated by the $\rho$. A
further simplification stems from considering the spectral density of
$\rho$ as a simple pole located at its peak which in turn dictates
that the coupling of $\rho$ to the electromagnetic current is
$em_\rho^2/g$.

Under the assumption of VDM and $\rho$ saturation, the expressions for
the thermal dilepton rates from longitudinal and transverse $\rho$
modes are given, respectively, by~\cite{Gale} 
\begin{widetext}
\be
   E_+E_-\frac{dR}{d^3p_+d^3p_-}&=&\frac{1}{(2\pi)^6}
   \left(\frac{e^4}{g^2}\right)
   \left(\frac{m_\rho^4}{M^4}\right)
   \left[q^2-\frac{({\mathbf q}\cdot{\mathbf k})^2}{k^2}\right]
   \frac{-{\mbox {Im}}F}{(M^2-m_\rho^2-{\mbox {Re}}F)^2 + {\mbox
   {Im}}F^2}\left(\frac{1}{e^{\beta\omega_L}-1}\right),\nonumber\\
   E_+E_-\frac{dR}{d^3p_+d^3p_-}&=&\frac{1}{(2\pi)^6}
   \left(\frac{e^4}{g^2}\right)
   \left(\frac{m_\rho^4}{M^4}\right)
   \left[2M^2-q^2+\frac{({\mathbf q}\cdot{\mathbf k})^2}{k^2}\right]
   \frac{-{\mbox {Im}}G}{(M^2-m_\rho^2-{\mbox {Re}}G)^2 + {\mbox
   {Im}}G^2}\left(\frac{1}{e^{\beta\omega_T}-1}\right),
   \label{dileptonrate}
\ee
\end{widetext}
where $P^\mu_+=(E_+,{\mathbf p}_+)$ and $P^\mu_-=(E_-,{\mathbf p}_-)$
are the positron and electron four momenta, respectively,
$K^\mu=P^\mu_++P^\mu_-$, $Q^\mu=P^\mu_+-P^\mu_-$, $M^2=k_0^2-k^2$ is
the pair invariant mass, $\omega_{L,T}$ are the longitudinal and
transverse $\rho$ modes dispersion relations and we have neglected the
electron mass. 

Figures~7 and~8 show the dilepton production rates as functions of
the pair invariant mass $M$ for ${\mathbf q}\cdot{\mathbf k}=0$ for
fixed values of $T=150$ MeV and $k$ for two values of $\mu=0, 100$ MeV
for $(a)$ longitudinal and $(b)$ transverse modes. Figure~7 considers
an small value of $k=50$ MeV and Fig.~8 a larger value $k=250$ MeV. In
both cases we can see that the effect of the finite chemical potential is
to moderately widen the distribution and to (also moderately) displace
the position of the peak toward larger values of $M$. The most
significant effect is however the lowering of the peak for finite $\mu$
which is in agreement with the analysis of Ref.~\cite{Rapp2}.   
    
\section{Summary and conclusions}\label{secVI}

In this paper we have considered the effects of a finite pion density
on the propagation properties of $\rho$ mesons at finite
temperature. The $\rho$ has been introduced by gauging an effective
Lagrangian obtained from the linear sigma model in the kinematical
regime where the pion mass and temperature are small compared to the
sigma mass. The finite density is described in terms of a finite pion
chemical potential associated to a conserved pion number. We have argued
that such description is important for ultra relativistic heavy-ion
collisions at RHIC and LCH energies where the central rapidity region
is expected to become baryon free. In this situations, we expect that
the influence of the dropping $\rho$ mass or melting or resonances
scenarios to describe the dilepton spectra, which rely on the
effects of a finite baryon density, become less important than the
effects of the expected large pion density. This is so particularly
during the hadronic phase of the reaction, between chemical and
thermal freeze-out when the strong interaction drives pion-number to a
fixed value.  

We have found that the $\rho$ thermal mass increases monotonically
with both the temperature and the pion density. However, the $\rho$
width as a function of $\mu$ and for fixed $T$ starts off by decreasing,
reaching a minimum at a finite value of $\mu$. This behavior can be
understood by noticing that as the density increases, the pion mass
does too and the phase space available for the decay process
$\rho\rightarrow \pi \pi$ narrows up to a --temperature dependent-- value
of $\mu$. From this value, the increase in the thermal $\rho$ mass is
stronger than the increase in the thermal pion mass and this effect
produces a widening of the phase space for the process, thus producing
the $\rho$ width to rise.  

Under the assumption of VDM and $\rho$ saturation, we have also
computed the dilepton production rate as a function of the pair
invariant mass. We have found that the finite pion density produces
a moderate broadening of the distribution and a moderate increase of
the position of the peak. The finite pion density
also produces a decrease of the distribution at the position of the
peak compared to the $\mu=0$ case.

The problem posed by the renormalization of theories
involving resummation is by no means a simple one. In fact, in recent
years this subject has been much actively pursued in the
literature (see for example Refs.~\cite{Hees, Caldas, Blaizot}). The
consensus is that when the theory is vacuum renormalizable, it 
will still be renormalizable after resummation. The solution, which
can be formulated using different languages, has been shown to require
that the counterterms needed for vacuum renormalization in ordinary
perturbation theory need also to receive the benefits of resummation
and be considered self-consistently in the resummation procedure.

For the purposes of our work, let us stress that the resummation we
have implemented for the pion self-energy corresponds to the tadpole
approximation (see Sec III of Ref.~\cite{Hees}). In fact, the gap
equation that renders the finite temperature and density corrections
to the pion mass in our work, Eq.~(\ref{solpi0}), is 
identical to the thermal part of Eq. (24) of the above reference; we
have just gone one step further, expanding the integrand as a series
which allows us to analytically integrate each term. In Eq. (24) of
Ref.~\cite{Hees}, the renormalization of the vacuum self energy and
four-point function have been carried out on the physical mass-shell
condition and have taken into account the self-consistency. In the
language of this reference, this is so because in the tadpole
approximation, the renormalized four-point function is constant and
given just in terms of the original coupling constant on the
mass-shell condition. 

Had we gone beyond the tadpole approximation, the situation would
have not be that straightforward. In fact, it has also been shown in
Ref.~\cite{Hees} that the self-consistent resummation scheme requires
a more complicated renormalization of the four-point function to be
taken into account alongside the renormalization procedure for the
self-energy. The effects of this certainly more complete analysis
might be interesting with regard to further thermal modifications to
the pion mass in situations where the original coupling constant of
the $\phi^4$ theory is much larger that one but is certainly beyond the
scope of the present work where our coupling constant $\alpha$, as
determined by the scale of the interactions, set by the vacuum pion
mass, is of order 1 (in fact, what matters, as shown in
Eq.~(\ref{solpi0}) is the effective coupling constant given by
$\alpha/(4\pi^2)$ which is of order 0.1). 

Finally, in order to predict the final dilepton spectra at RHIC and LHC
energies, the results found in this work have to be placed into a
model for the evolution of the collision and also possibly to include
the effects of other vector or axial vector mesons. All this is for
future. 
 
\section*{Acknowledgments}

A. Ayala wishes to thank Centro Brasileiro de Pes\-qui\-sas Fi\-si\-cas for
their kind hospitality during the time when part of this work was done.
Support for this work has been received in part by DGAPA-UNAM under PAPIIT
grant number IN108001 and CONACyT under grant numbers 32279-E and 40025-F.

\end{document}